%% file: miles_params.tex
\documentclass{aa}
\usepackage{graphicx,color}
\usepackage{natbib}
\usepackage{url}
\usepackage{amsmath}
\usepackage{amssymb}
\usepackage{longtable}
\usepackage{multirow}

\bibpunct{(}{)}{;}{a}{}{,} 
\newcommand{\ulyss}{ULySS}

\newcommand{\kms}{km\,s$^{-1}$}
\begin{document}
 \title{The atmospheric parameters and spectral interpolator for the stars of MILES}

  \author{  Philippe Prugniel,
          \inst{1}
 \and
 Isabelle Vauglin
 \inst{1}
 \and
 Mina Koleva
 \inst{2,1}
        }

   \offprints{Ph. Prugniel}

   \institute{Universit\'e de Lyon, Universit\'e Lyon 1, Villeurbanne, F-69622, France;
              CRAL, Observatoire de Lyon, CNRS UMR~5574, 69561 Saint-Genis Laval, France\\
              \email{philippe.prugniel@univ-lyon1.fr}
              \email{isabelle.vauglin@univ-lyon1.fr}
              \and
              Instituto de Astrof\'{\i}sica de Canarias, La Laguna, E-38200 Tenerife, Spain; Departamento de Astrof\'{\i}sica, Universidad de La Laguna, E-38205 La
Laguna, Tenerife, Spain\\
              \email{koleva@iac.es}
 }

   \date{Received ; accepted }


  \abstract
  { 
Empirical libraries of stellar spectra are used for
stellar classification and synthesis of stellar populations.
MILES is a medium spectral-resolution library in the optical domain
covering a wide range of temperatures, surface gravities and metallicities.
}
  {
We re-determine the atmospheric parameters of these stars in order
to improve the homogeneity and accuracy. We build an
interpolating function that returns a spectrum as a function
of the three atmospheric parameters, and finally, we characterize
the precision of the wavelength calibration and stability of the
spectral resolution.
}
    {
We use the \ulyss{} program with the ELODIE library as a reference and
compare the results with literature compilations.
     }
     {
We obtain precisions of  
60 K, 0.13 and 0.05 dex 
respectively for $T_{\rm{eff}}$, log g and [Fe/H] for the FGK stars. 
For the M stars, the mean errors are 38~K, 0.26 and 0.12 dex,
and for the OBA 3.5\%, 0.17 and 0.13 dex.
We construct an interpolator that we test against the MILES
stars themselves. We test it also by measuring the
atmospheric parameters of the CFLIB stars with MILES as reference
and find it to be
more reliable than the ELODIE interpolator for the evolved hot stars, like
in particular those of the blue horizontal branch.
}
   {}
 \authorrunning{Prugniel et al.}
   \titlerunning{Stellar atmospheric parameters for MILES}

  \maketitle
%

\section{Introduction}

MILES \citep{miles}
is a medium resolution library of observed stellar spectra in the
optical domain.
It is comparable to CFLIB \citep{valdes2004} and ELODIE
\citep{ps2001}, and is of a particular interest for its accurate 
flux calibration.
The three libraries contain {\it normal} stars
with a wide range of characteristics, from spectral types O to M, 
all luminosity classes and a wide distribution of metallicities 
($ -2.5 <$ [Fe/H] $ < 1 $) dex.

The empirical libraries have important applications in different fields.
They are used as references for stellar classification and
determination of atmospheric parameters
\citep[see][and references therein]{wu2011}. 
They are also important ingredients for the synthesis of stellar
populations, used to study the history of galaxies
\citep{prugniel2007}.
The most important characteristics of a library are
(i) the wavelength range, (ii) the spectral resolution
and (iii) the distribution of the stars in the parameters' space
whose axes are the effective temperature, $T_{\rm{eff}}$, the
logarithm of the surface gravity, log~g, and the metallicity, [Fe/H].
Other properties, like the precision and uniformity of the
wavelength calibration and spectral resolution, or the
accuracy of the flux calibration, are also to be considered.

The ELODIE library has been upgraded three times after its publication
\citep{ps2001,elo30,elo31}. The last version, ELODIE 3.2 is preliminary 
described in \citet{wu2011}.
It counts 1962 spectra of 1388 stars observed with the eponym
echelle spectrograph \citep{baranne1996} at the spectral resolution
$\Delta\lambda \approx 0.13$ \AA{} 
(R=$\lambda/\Delta\lambda \approx 42000$), in the wavelength range 
3900 to 6800 \AA.
CFLIB, also known as the ``Indo-US'' library, has 1273 stars
at a resolution\footnote{
\citet{valdes2004} mention  $\Delta\lambda \approx$ 1.2~\AA{},
but \citet{beifiori2010} and \citet{wu2011} derive 1.4~\AA{}.}
$\Delta\lambda \approx$ 1.4~\AA{} ($3000 \lessapprox$ R $\lessapprox 6000$) 
in the range 3460 to 9464~\AA.
The atmospheric parameters of CFLIB were homogeneously
determined by \citet{wu2011}.
MILES contains 985 stars at a resolution\footnote{
The value  $\Delta\lambda =$ 2.3~\AA{} \citep{miles} is underestimated, see
\citet{beifiori2010} and Sect.~\ref{sect:result:lsf}}
$\Delta\lambda \approx 2.56$ \AA{} in the range 3536 to 7410~\AA. 
The atmospheric parameters
of these stars were compiled from the literature or derived
from photometric calibrations by \citet{cenarro2007}.
The [Mg/Fe] relative abundances were recently determined by
\citet{decastromilone2011}.

The goals of this article are to (i) re-determine homogeneously 
the atmospheric parameters of the stars of MILES using ELODIE as reference, 
(ii) characterize the resolution and accuracy of the wavelength calibration
and (iii) build an {\it interpolator}.
This latter is a function, based on an interpolation over all the
stars of the library, that returns a spectrum for a given set of atmospheric
parameters , $T_{\rm{eff}}$, log~$g$, and  [Fe/H].

In Sect.~\ref{sect:anal}, we describe the steps of the data analysis.
In Sect.~\ref{sect:results}, we present the results and assess their
reliability, and Sect.~\ref{sect:conclusion} gives the conclusions.

\section{Analysis}
\label{sect:anal}

In this section, we give the details of our analysis. First, we describe
the different steps, and then we present in details the
determination of the atmospheric parameters and line-spread function
and the computation of the interpolator.

\subsection{Strategy}

To determine the atmospheric parameters, we compare the observed MILES
spectra with templates built from the ELODIE library. The $\chi^2$
minimization, performed with the ULySS 
program\footnote{\url{http://ulyss.univ-lyon1.fr}}
\citep{ulyss}, is made as described in \citet{wu2011}.
Shortly, the underlying model is
\begin{align}
{\rm Obs}(\lambda) = P_n(\lambda)  \times {\rm G}~\otimes 
\rm{TGM}(T_{\rm eff}, g, \mathrm{[Fe/H]}, \lambda),
\label{eqn:main}
\end{align}

where $\rm Obs(\lambda)$ is the observed spectrum sampled in log$\lambda$, 
$\rm P_{n}(\lambda)$ a series of Legendre polynomials of degree n, and $\rm
G(v_{res},\sigma)$ a Gaussian broadening function parameterized
by the residual velocity $\rm v_{res}$, and the dispersion $\sigma$. 
The TGM function models a stellar spectrum for given atmospheric 
parameters. It interpolates the ELODIE 3.2 library described
in Sect.~\ref{sect:elointerp}.
The program minimizes the squared difference between the observations and 
the model. 
The free parameters are the three of TGM, the two  of G and the n 
coefficients of $\rm P_{n}$. 

A single minimization provides the atmospheric parameters and the broadening.
The advantage of this simultaneity is to reduce the effects of the
degeneracy between the broadening and the atmospheric parameters
\citep[see][]{wu2011}.

The function G encompasses both the effects of the finite spectral resolution 
and of the physical broadening of the observation and model. The physical
broadening is essentially due to rotation and turbulence. The
spectral resolution is represented by the so-called line-spread function (LSF), and
in first approximation we can write:
\begin{align}
G =\mathrm{LSF}_{rel} \otimes \Phi
\label{eqn:broad}
\end{align}
where $\Phi$ is the relative physical broadening between the observation
and the model (i.~e. mismatch of rotation and turbulence) and 
$\mathrm{LSF}_{rel}$ the relative LSF. 
The absolute LSF of the observed spectrum is $\mathrm{LSF} = \mathrm{LSF}_{mod} \otimes \mathrm{LSF}_{rel}$, where   $\mathrm{LSF}_{mod}$ is the LSF of the model.
The approximations are that (i) neither $\Phi$ nor $\mathrm{LSF}_{rel}$ are strictly Gaussians, 
and (ii) the LSF generally depends on the wavelength, hence we cannot rigorously
write convolutions.
The Gaussian approximation is certainly acceptable in the present context
of moderate spectral resolution because: 
(i) The physical broadening can often be neglected or can otherwise be assumed 
Gaussian. 
(ii) The MILES spectra were
acquired with a relatively narrow slit, thus the top-hat signature of the 
slit is dominated by the intrinsic broadening due to the disperser.

The variation of the LSF with the wavelength has only minor consequences on the
atmospheric parameters \citep[see][]{wu2011}, but we will explain below
how we determine it and inject it in TGM to get the most accurate parameters.

In Eq.~\ref{eqn:main}, the role of the multiplicative polynomial, 
$\rm P_{n}(\lambda)$, is to absorb the mismatch of the shape of
the continuum, due to uncertainties in the flux calibration. 
It does not bias the measured atmospheric parameters,
because it is included in the fitted model rather than determined
in a preliminary {\it normalization}. 
In principle, a moderate degree, n~$\approx 10$,
is sufficient, but a higher degree suppresses the `waves' in the residuals
and helps the interpretation of the misfits (the residuals
are smaller and it is easier to detect poorly fitted lines).
Large values of n, up to 100 or more, do not affect the parameters
\citep{wu2011}.
The optimal choice of n depends on the resolution, wavelength range and
accuracy of the wavelength calibration.
We determined it following the precepts of \citet{ulyss}. We chose
stars of various spectral types, and tested different values in order
to locate the plateau where the atmospheric parameters are not sensitive 
to n.
We adopted n~$=~40$.

The choice of ELODIE as reference limits the wavelength range where the 
spectra can be analysed. In particular, the blue end, below the H \& K lines,
is unfortunately not used. An alternative would have been to use a theoretical
library, like the one of \citet{coelho2005}. 
We tried this solution, but we found that the misfits
are significantly larger than with ELODIE  
(see Sect.~\ref{sect:result:lsf})
and we decided to maintain our initial choice.

In order to handle the wavelength dependence of the LSF, the analysis
proceeds in three steps:

\begin{description}

\item[{\it Determination of the LSF}.]
We determine the wavelength-dependent LSF of each spectrum
of stars in common between the MILES and ELODIE libraries.
We use the \textsc{uly\_lsf} command, as described in
Sect.~\ref{sect:lsf}.

\item[{\it Determination of the atmospheric parameters.}]
We inject the wavelength-dependent relative LSF into the models
so that the result has the same resolution characteristics
as the observations,
and determine the atmospheric parameters calling
\textsc{ulyss}.

\item[{\it Construction of the spectral interpolator.}]
Finally, using these atmospheric parameters, we compute an interpolator.
For each wavelength element, a polynomial in  
$\log T_{\rm{eff}}$, log~$g$, and  [Fe/H] is adjusted on all the library
stars, in order to be used as an interpolating function.
This process is introduced in Sect.~\ref{sect:inter}.

\end{description}

\subsection{ELODIE 3.2: library and interpolator}
\label{sect:elointerp}

ELODIE 3.2 is based on the same set of stars as
ELODIE 3.1 \citep{elo31} and benefited from several improvements
concerning various details of the data reduction, like in particular
a better correction of the diffuse light. 
We note also that a systematic error of 0.0333 \AA{} (i.~e. approximately 2 \kms)
on the wavelengths of the previous version has been corrected
(it was due to a bug in the computation of the world coordinate system
after a rebinning; ELODIE 3.1 was red-shifted).

The ELODIE interpolator approximates each spectral
bin with polynomials in  $T_{\rm{eff}}$, log~g and [Fe/H]. Three different
sets of polynomials are defined for the OBA, FGK and M type temperature
ranges, and are linearly interpolated in overlapping regions.
This interpolator has been
noticeably upgraded in the last version, taking into account the stellar 
rotation and
adding some theoretical spectra to extend its range of validity to
regions of the parameters' space scarcely populated of devoid of library stars.
The ELODIE 3.2 interpolator is publicly available at
\url{http://ulyss.univ-lyon1.fr/models.html}, and additional details
are given in \citet{wu2011}.
A similar interpolator is described in Sect.~\ref{sect:inter} for MILES,
with the only difference that the rotation terms are omitted.

\subsection{Accurate line-spread function}
\label{sect:lsf}

The LSF describes the instrumental broadening,
and may vary with the wavelength.
We determined the wavelength dependent broadening
by fitting the spectra of the 303 MILES stars belonging also to
ELODIE 
(note that since ELODIE contains repeated observations of the same stars, 
this corresponds to 404 comparisons).
These fits were performed with
the function \textsc{uly\_lsf} in a series of 400 \AA{} intervals separated
by 300 \AA, therefore overlapping by 100 \AA{} on both ends.
This procedure gives nine sampling points along the ELODIE range.

The change of broadening with wavelength is a consequence of the characteristics
of the disperser and design of the spectrograph, but the shift of these functions
with respect to the rest-frame wavelengths shall ideally be null. 
However, the finite precision of the wavelength calibration 
and uncertain knowledge of the heliocentric velocity of the
stars result in residual shifts that may be wavelength dependent.  
Flexures in the spectrographs or temperature drifts may cause these effects.
Their magnitudes are expected to be small fractions of  
pixels. \citet{miles} estimated the precision of their calibration to about
6~\kms.These residuals are likely to cancel each others when
we average the LSF for all the stars.

We estimated the mean instrumental velocity dispersion and residual shift
in each spectral chunk 
as a clipped average of the individual ones using the IDL
procedure \textsc{biweight\_mean} that does a bisquare weighting
(a median estimation gives identical results).

\subsubsection{Absolute LSF}

Our analysis is providing the relative LSF between MILES and ELODIE.
Since the characterization of the LSF has an intrinsic interest, we
will give the absolute LSF obtained deconvolving by the LSF of ELODIE.

The FWHM resolution of the ELODIE spectrograph, measured on the Thorium 
lines of 
calibrating spectra varies from 7.0 \kms{} in the blue to 7.4 in the red
\citep{baranne1996}, or respectively 0.09 and 0.17 \AA.
This corresponds to a mean resolving power of R$ = 42\,000$.
The low-resolution (i.~e. R~$ \approx$~10\,000) version of ELODIE 3.2, 
used in this paper, was produced by
convolving the full-resolution spectra with a Gaussian of FWHM = 0.556~\AA.
Therefore, the final resolution varies from 0.564 to 0.581~\AA{} along the
wavelength range, for an average of 0.573~\AA{}.

To check this value, we analysed the LSF
of the ELODIE interpolated spectra, having the atmospheric parameters of the 
MILES 
stars, using \citet{coelho2005} as reference.
We found a relative broadening of $0.584 \pm 0.006$\AA, independently of the
wavelength. The difference with the value above is surely compatible
with the residual rotational broadening of the interpolated spectra,
and we adopt the mean value derived above.

The Gaussian width of absolute LSF of MILES is therefore the quadratic sum
between the width of the LSF relative to ELODIE and the width of the absolute
LSF of ELODIE.

\subsubsection{Biased LSF}

If the effective spectral resolution was the same for all the spectra of MILES,
we could simply inject the LSF into the model and adjust
only the atmospheric parameters 
(i.~e. omit the convolution in Eq.~\ref{eqn:main}). 
However, because of the rotational
broadening and dispersion of the instrumental broadening,
the effective resolution varies, and
we still need to fit the atmospheric parameters {\it and} the broadening.

In practice, the model must have a
higher spectral resolution than the observation, 
because it
is convolved with G during the analysis (Eq.~\ref{eqn:main}).
If we would inject the relative LSF in the model,
the result would be broader than the best resolved library spectra.
To avoid this difficulty we bias
the LSF by subtracting quadratically 40~\kms (at any wavelengths)
from the width of the mean relative LSF. 
The resolution of this biased LSF is higher
than any spectrum of the library, and it has the correct wavelength dependence.

\subsection{Determination of the atmospheric parameters}
\label{sect:tgm}

We fitted the spectra using the ELODIE 3.2 interpolator, injecting
the biased LSF previously derived, and assuming a
uniform broadening, as described in Eq.~\ref{eqn:main}.
In order to avoid trapping in local minimal, we used a grid of initial guesses
sampling all the parameters' space. The nodes of this grid are:
\begin{itemize}
\item[]$T_{\rm{eff}} \in \{3500, 4000, 5600, 7000, 10000, 18000, 30000\}$~K
\item[]log g~$ \in \{1.8, 3.8\} $~cm\,s$^{-2}$ 
\item[][Fe/H]~$ \in \{-1.7, -0.3, 0.5\}$
\end{itemize}
For the stars belonging to clusters, we adopted and fixed the metallicity 
to the value given in \citet{cenarro2007}.

The spectra were rebinned into an array of logarithmically
spaced wavelengths, each pixel corresponding to 30~\kms{}.
This choice oversamples the original spectrum by a factor two in the blue
and by 20\% in the red. 
We performed the fit in the region 4200 -- 6800 \AA, excluding the blue
end of the spectra, where the signal-to-noise ratio is lower.

Because the noise estimation in the MILES spectra is not available,
we assumed a constant noise, resulting in a uniform weighting of each
wavelength bin. We estimated an upper limit to the internal errors on the 
derived parameters by
assuming $\chi^2 = 1$.

This first minimization localizes the region of the solution, and we
refine our measurements 
running again \textsc{ulyss} with the \textsc{/clean} option to identify
and discard the spikes in the signal. They result from the imperfect
subtraction of sky lines, removal of spikes due to hits of cosmic rays
or stellar emission lines. The second set of derived parameters is very close 
to the first one, because the MILES spectra were already corrected for most of 
the observational artifacts. 

Finally, the resulting parameters were compared with \citet{cenarro2007}
and the significant outliers were examined by checking the quality
of the fit and searching the literature for accurate measurements from high 
resolution spectroscopy.

\subsection{MILES library interpolator}
\label{sect:inter}

The goal is to build an interpolator similar to the one of the ELODIE library.
It may then be used to (i) analyse stellar spectra, for example with ULySS,
or (ii) create stellar population models, for example with PEGASE.HR
\citep{leborgne2004}.

The general idea is to approximate each wavelength bin with a
polynomial function of  $T_{\rm{eff}}$, log g and [Fe/H].
This process resembles to the {\it fitting functions} \citep{worthey1994}
that are used to predict the equivalent width of some features, or 
spectrophotometric indices given some atmospheric parameters. It is 
extended to model every spectral point.

This is a {\it global} interpolation, in the sense that each polynomial is
valid in a wide range of parameters. An alternative would be to use a 
{\it local} interpolation, like averaging the nearest spectra to a given
point in the parameters' space. A good example of local interpolation 
is \citet{vazdekis2003}. Both methods have their own advantages and 
inconvenients.
The global interpolation is less sensitive to the stochasticity of the
distribution of the stars, but may not respond accurately in the regions
where the spectrum changes rapidly. It is also continuous and derivable 
everywhere, which are required properties to use it as a function for
non-linear fit, as in ULySS.
In both cases it is possible to 
control the quality of the interpolation by comparing each star to the
interpolated spectrum which match its parameters.

For the present work, we use the same polynomial developments as for
ELODIE 3.2, because this will permit to use it directly as a model for
a TGM component in ULySS.
The first version of this interpolator was described in \citet{ps2001},
and we remind below the principles and present the difference
introduced in ELODIE 3.2.

\subsubsection{ $T_{\rm{eff}}$ regimes}

The library contains all types of stars, from O to M, and the temperature
is the main parameter controlling the shape of the spectra. Modelling
all the stars with a single set of polynomials would necessitate to
include a large number of terms. The result would accordingly be very
unstable, presenting oscillations and violently diverging near the
edges of the parameters' space. 
For this reason, we defined three temperature ranges, matching 
the OBA, FGK and M spectral types, where independent set of polynomials
are adjusted.
These three regimes have comfortable overlaps, allowing us to connect them
smoothly by a linear interpolation. The limits are:
\begin{description}
\item[OBA regime:] $T_{\rm{eff}} > 7000 $~K
\item[FGK regime:] $ 4000 < T_{\rm{eff}} < 9000$~K
\item[M regime:] $T_{\rm{eff}} < 4550 $~K
\end{description}

Note that the M regime encompasses the cool K-type stars.

\subsubsection{Polynomial developments}

The developments are the same as for ELODIE.3.2, but truncated to
exclude the rotation terms introduced to suppress a bias due to
a degeneracy between the stellar rotation and the temperature
\cite[see][]{wu2011}. Because of the lower spectral resolution of MILES,
the stellar rotation is mixed with the variation of the resolution
from star to star, and
the introduction of these terms did not appear relevant.

The terms were chosen iteratively, adding at each step the one leading to
the largest reduction of the residuals between the observations and the 
interpolated spectra.
The following developments were used:

\begin{align}
{\rm TGM}&(T_{\rm eff}, g, \mathrm{[Fe/H]}, \lambda) = \nonumber\\
&a_0(\lambda) +  a_1(\lambda) \times \log T_{\rm eff} +
a_2(\lambda) \times [Fe/H] + \nonumber\\
&a_3(\lambda) \times  \log g + 
a_4(\lambda) \times (\log T_{\rm eff})^2 + \nonumber\\
&a_5(\lambda) \times (\log T_{\rm eff})^3 +
a_6(\lambda) \times (\log T_{\rm eff})^4 + \nonumber\\
&a_7(\lambda) \times  \log T_{\rm eff} \times [Fe/H] +
a_8(\lambda) \times  \log T_{\rm eff} \times  \log g +  \nonumber\\
&a_9(\lambda) \times (\log T_{\rm eff})^2 \times  \log g + \nonumber\\
&a_{10}(\lambda) \times (\log T_{\rm eff})^2 \times  [Fe/H] + \nonumber\\
&a_{11}(\lambda) \times (\log g)^2 +
a_{12}(\lambda) \times ([Fe/H])^2 + \nonumber\\
&a_{13}(\lambda) \times  (\log T_{\rm eff})^5 +
a_{14}(\lambda) \times \log T_{\rm eff} \times (\log g)^2 +\nonumber\\
&a_{15}(\lambda) \times  (\log g)^3 +
a_{16}(\lambda) \times ([Fe/H])^2 + \nonumber\\
&a_{17}(\lambda) \times \log T_{\rm eff}  \times  ([Fe/H])^2 +  \nonumber\\
&a_{18}(\lambda) \times \log g \times [Fe/H] +\nonumber\\
&a_{19}(\lambda) \times (\log g)^2 \times [Fe/H] +\nonumber\\
&a_{20}(\lambda) \times  \log g  \times ([Fe/H])^2 +\nonumber\\
&a_{21}(\lambda) \times T_{\rm eff} + 
a_{22}(\lambda) \times  (T_{\rm eff})^2 
\label{eqn:dev}
\end{align}

TGM is a flux-calibrated interpolated spectrum. Unlike for ELODIE, we
did not compute a continuum-normalized interpolator, as it is not needed here.

The 23 terms were used for both the FGK and M regimes, but the development
was truncated to the first 19 for the OBA one. 

\subsubsection{Support for extrapolation}

One of the limitation of using empirical libraries is that they do not cover
all the range of atmospheric parameters we may wish they would. 
In particular, to study the stellar populations of galaxies, we would need,
for example, young stars of low metallicity, which are obviously missing
in a library of Galactic stars.

For this reason, it is important that the interpolator preserves its quality
at the edges of the parameters' space, where only rare stars are present.
This is a difficulty for any type of interpolating function.

A solution could have been to supplement the library with theoretical spectra
in the margins of the parameters' space. However, this would introduce
discontinuities because the flux scale of theoretical spectra is not
fully consistent with the empirical library.
To improve this situation, \citet{prugniel2007} introduced a semi-empirical solution where 
theoretical spectra are used {\it differentially} to extend the coverage of the parameters' space.
This was used to add a [Mg/Fe] dimension to the space,
and to model spectra with non-solar abundances, as Galactic
globular clusters \citep{prugniel2007,koleva2008}.
The same principle was adopted in ELODIE 3.2 to extend the range of the 
3-dimensional parameters' space (without the [Mg/Fe] dimension which
is not taken into consideration neither in ELODIE 3.2 nor in the present paper).

We compute the differential effect of changing
a parameter between a point belonging to the empirical library, and another
one located outside of the range of the library. This differential spectrum
is built using a theoretical library. Finally, we produce 
a semi-empirical spectrum, 
summing the differential one to one generated with the initial version of
then interpolator 
(computed without the semi-empirical extrapolation supports) 
at the reference location.

We used the \citet{martins2005} library to add semi-empirical spectra at the
following locations:
(i) $T_{\rm{eff}} = 40000$~K, log~$g =$~4 and 4.75, and [Fe/H]~$=~-1$, 0 and +0.3 dex,
using as reference $T_{\rm{eff}} = $ 20000 K, log~$g =$~3.5, [Fe/H]~$=$0;
(ii) $T_{\rm{eff}} = 55000$~K, log~$g =$~3.5 and 4.75, and [Fe/H]~$=~-1$, and +0.3 dex,
using as reference $T_{\rm{eff}} = $ 30000 K, log~$g =$~3.5, [Fe/H]~$=$0;
(iii) $T_{\rm{eff}} = 20000$~K, log~$g =$~3, and 5 and [Fe/H]~$=~-1$,
using as reference $T_{\rm{eff}} = $ 20000 K, log~$g =$~3.5, [Fe/H]~$=$0.
We also used the \citet{coelho2005} library to add some low metallicity cool
dwarfs at the locations $T_{\rm{eff}} = $ 3500 K, log~$g =$~4.5 and 5.0, [Fe/H]~$= -1.5, -2.0$ and $-2.5$   using as reference  $T_{\rm{eff}} = $ 3500 K, log~$g =$~4.5, [Fe/H]~$= -0.5$.

We affect a low weight to these
spectra, and they do not affect the region populated with observed stars:
Each extrapolation-support spectrum has 1/20th of the weight of an observed star.
We computed a final version of the interpolator, using the
semi-empirical spectra to prevent the divergence at the edges of the parameters'
space and to extend the validity range. 
The interpolated spectra in the extrapolated regions are probably not very 
accurate, but they do not diverge and are sufficient for many applications.

\section{Results}
\label{sect:results}

In this section, we present the results of the previous procedure.
We discuss the determination of the LSF, 
the measurements of the atmospheric parameters,
and finally the computation and
validation of the interpolator.

\begin{figure}
\includegraphics[width=9cm]{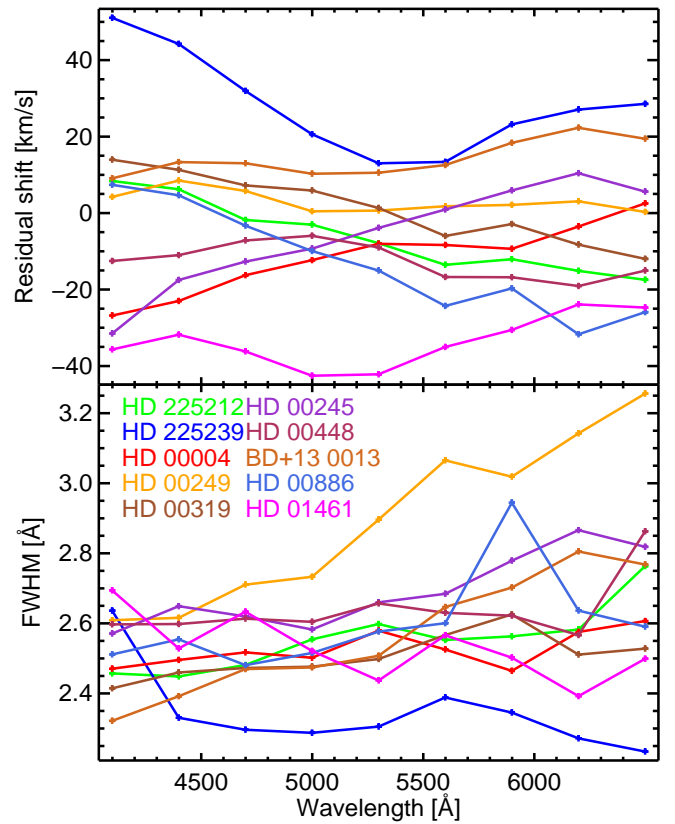} 
\caption {
Line-spread function for 10 stars of the MILES library chosen arbitrarily (actually 10 of the first 12), using for reference the interpolated ELODIE spectra.
The top panel shows the residual shift of the spectra, illustrating the precision
of the wavelength calibration and of the rest-frame reduction. 
The bottom panel presents the FWHM resolution. 
The mean formal error on each LSF point is of 0.5~\kms{} on the residual shift, and
0.025~\AA{} on the FWHM resolution
}
\label{fig:indivlsf}
\end{figure}
\begin{figure}
\includegraphics[width=9cm]{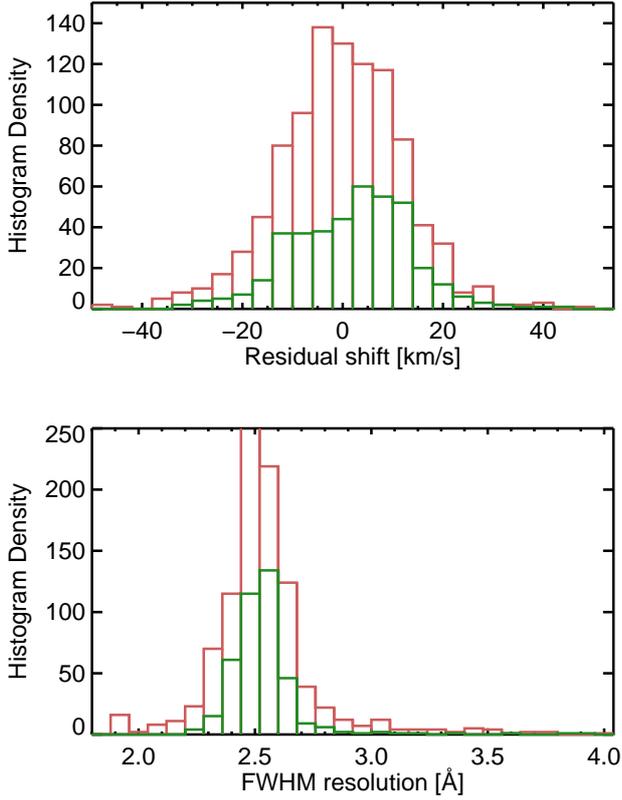} 
\caption {
Histograms of the broadening and residual shift of the line-spread function of the MILES library at 5300~\AA. 
The green histograms are for the 404 direct comparisons with spectra of the ELODIE library.
The red ones are the comparisons with the ELODIE interpolator.
The top panel is the distribution of the residual shifts, in \kms, and the
bottom ones the distribution of the FWHM Gaussian broadening.
}
\label{fig:histolsf}
\end{figure}

\subsection{Line-spread function and wavelength calibration}
\label{sect:result:lsf}

The broadening was determined individually by comparing MILES and ELODIE
spectra for all the stars in common.
In order to increase the statistics, we also made the analysis for all
the MILES stars by comparing them with the ELODIE interpolated spectrum
corresponding to their atmospheric parameters. 
This second set of LSFs includes both the instrumental
and physical broadening of the individual stars.

Figure~\ref{fig:indivlsf} presents the individual LSF (using the 
ELODIE interpolated spectra as reference)
for some stars chosen arbitrarily (the firsts of the list).
From this small subset alone, it is apparent that the broadening
is variable. Some spectra have a lower effective resolution, possibly
due to stellar rotation, and some have a higher resolution, maybe because of
a better focusing of the spectrograph.
It also appears that the rest-frame reduction is not always accurate, with
deviation reaching a few 10 \kms. This may be due to (i) uncertain knowledge
of the heliocentric velocities, (ii) imperfect wavelength calibration, or
(iii) stellar duplicity. We note also that the residual shift often changes
with the wavelength by 10 to 30 \kms{} over the ELODIE range. This results
from an uncertainty in the dispersion relation. The effect is slightly 
larger that the precision estimated in \citet{miles}.
The values of the broadening and residual shift at 5300~\AA{} 
are given for each star in Table~\ref{table:params}.

The
histograms of the broadening and residual shifts are presented on Fig.~\ref{fig:histolsf}.
The gaussian broadening at 5300~\AA{}
spans the range $30 < \sigma_{ins} < 92 $ \kms, 
(i.~e. $1.3 < FWHM < 3.8$ \AA) and
the histogram is skewed toward the large dispersions. This is likely due
to the effect of the rotation. 
The mean broadening at the same wavelength is 60.5 \kms, for the direct comparison,
with the rms dispersion of 2.4 \kms 
(i.~e. respectively 2.52 and 0.10 \AA{} for the FWHM).
The mean broadening is similar (60.9 \kms) 
and the spread slightly larger (3.6 \kms), when interpolated spectra are used.
The consistency between the two determinations
shows that the physical broadening is only a minor contribution.

As expected, the residual shifts essentially cancel in the mean LSF. 
The mean shift is 2 \kms (identical for the two analysis), in the sense
that MILES is red-shifted. 
The internal rms spread of these residual shifts is 12 \kms{} or FWHM=0.50 \AA{}
at 5300~\AA. 
If MILES is
used to compute population models, this will be combined with the instrumental
broadening. In other words, the resolution of an interpolated MILES
spectrum, or of a population model, will be: 
$\rm FWHM_{inter}  = \sqrt{2.52^2 + 0.50^2} = 2.57$ \AA.
(assuming that the effect is uniform over all the parameters' space).

The mean absolute difference of the residual velocity between the last and first 
segment of the LSF (i.~e. between 6500 and 4100\AA) is 15 \kms{}. This reflects
the accuracy of the dispersion relation used for the wavelength calibration.
As explain in \citet{miles},
to save observing time, arc spectra were not acquired for each
individual spectrum, but only for some spectra representative of each 
spectral type and luminosity class. It was assumed that the linear dispersion 
and higher order terms of the dispersion relation were constant, and 
a global shift was determined by cross-correlating each spectrum to a 
well-calibrated one. Our present test indicates that the stability of
the spectrograph was slightly over-estimated, and the variation of the
linear term of the dispersion relation will contribute for a further 
degradation of the LSF for population models.

\begin{figure}
\includegraphics[width=9cm]{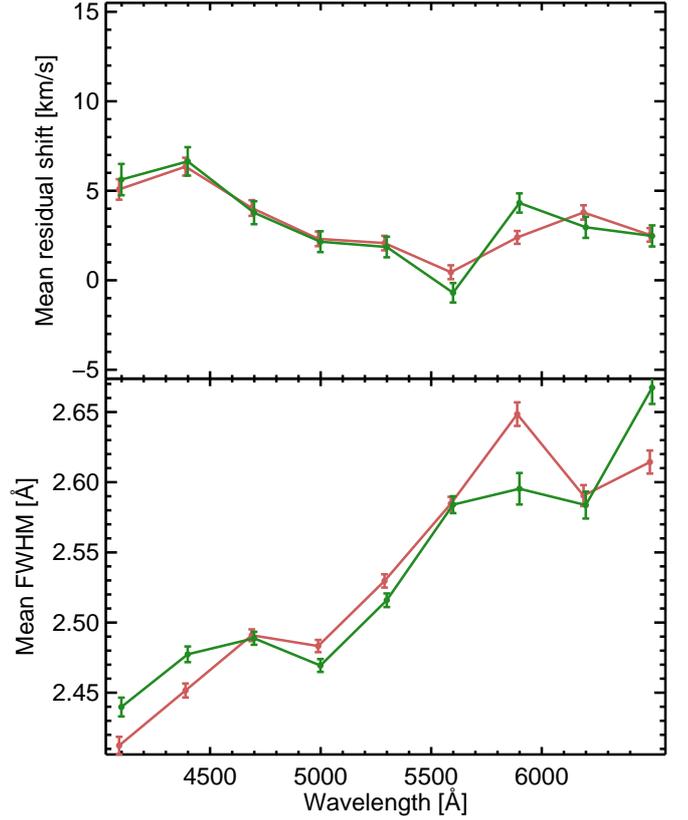} 
\caption {
Mean line-spread function of the MILES library, using for reference (i) the interpolated ELODIE spectra (red line and symbols) and (ii) the ELODIE spectra of stars in common (green). 
The top panel gives the mean residual shift over all the library, and the 
bottom panel the mean FWHM wavelength resolution. 
The bars indicate the errors on the mean value (i.~e. dispersion / $\sqrt{n-1}$).
The abscissa of the red symbols are shifted by a small quantity to avoid superposition.
}
\label{fig:meanlsf}
\end{figure}

The variation of the LSF with wavelength,
presented in Fig.~\ref{fig:meanlsf}, is consistent for the two sets of templates.
The resolution changes from 2.45 \AA{} at 4000 \AA, to 2.63 \AA{} at 6500 \AA,
with an average value of  2.56 \AA.
This is broader than the estimation in \citet{miles}.
In this paper, the authors found FWHM=2.3~\AA{} by comparison with CFLIB,
for which they assumed a resolution of 1 \AA. In fact, the resolution
of CFLIB is rather $\approx$  1.4 \AA{} \citep{wu2011, beifiori2010}, and
correcting this error put the two values in agreement.
\citet{beifiori2010} also measured the resolution of MILES with a similar
method and found 2.55~\AA, independent of the wavelength.
This is consistent with our result.

It is also interesting to characterize the LSF over the whole MILES
wavelength range. We therefore repeated the analysis using the 
\citet{coelho2005} library. We found consistent results in the ELODIE range,
but with a larger spread, certainly due to lower quality fits. The residuals
are typically three times larger than those obtained when we compared to 
ELODIE. A consequence is that the trend of the LSF with the wavelength is
smeared out, leaving a uniform $\rm FWHM = 2.59 \pm 0.08$~\AA{}.
We similarly analysed the MILES spectra of the five closest analogs of the Sun
against the high resolution spectrum from \cite{nsoa}. The results
are also consistent with those obtained with ELODIE, but with a large
spread resulting from the small statistics. 
Therefore we cannot constrain the resolution outside of the 
wavelength range of ELODIE with the same accuracy.

\begin{figure}
\includegraphics[width=9cm]{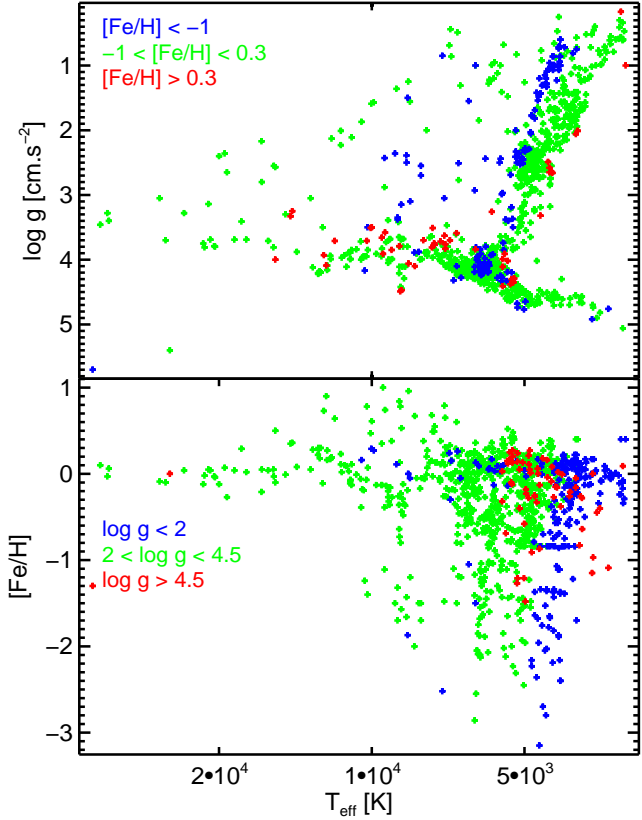} 
\caption {Distribution in the log($T_{\rm{eff}}$) - log~$g$ and
log($T_{\rm{eff}}$) - [Fe/H] planes of
the adopted atmospheric parameters for the 985 MILES stars. 
In the top panel, 
the color of the symbols distinguishes different metallicity classes.
In the bottom panel, it distinguishes different classes of surface gravity.
} \label{fig:distri_tg}
\end{figure}

\begin{figure*}
\centering
\includegraphics{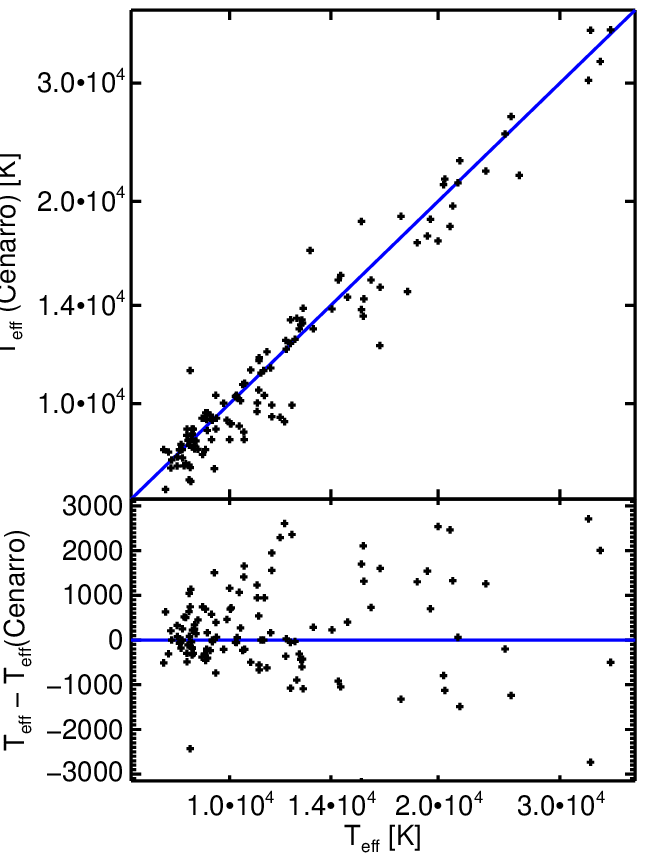} 
\includegraphics{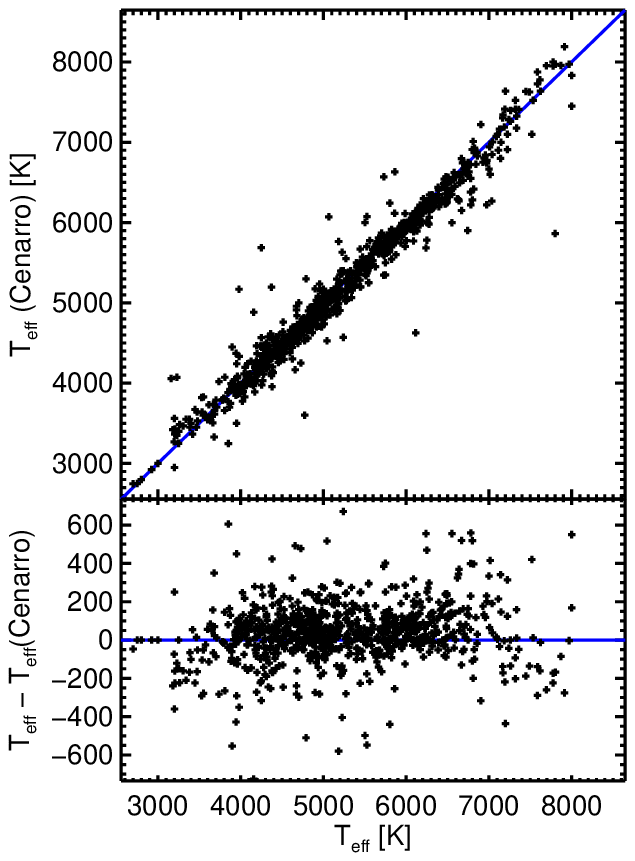} \\
\includegraphics{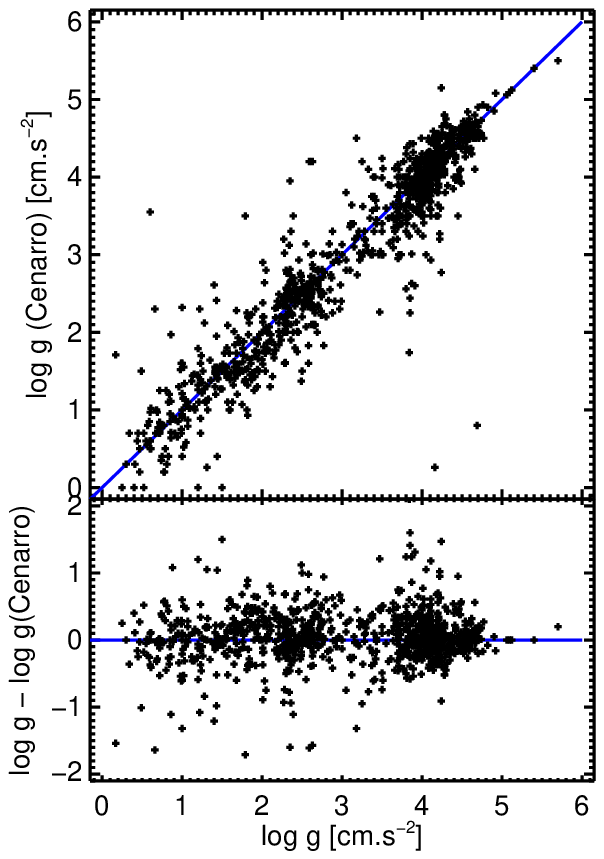} 
\includegraphics{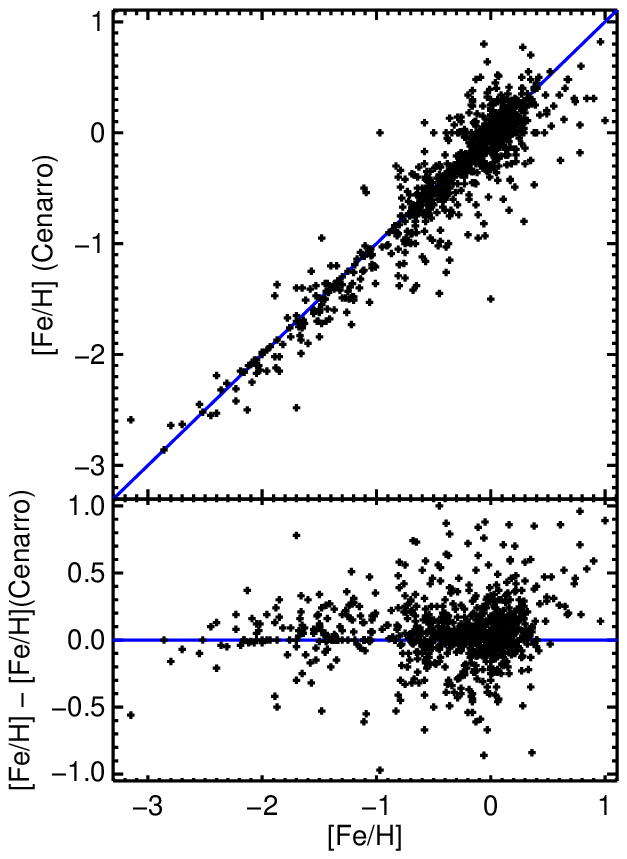} 
\caption {
Comparison of the measured atmospheric parameters with the \citet{cenarro2007} compilation. The abscissae are the parameters measured in the present paper.
}
\label{fig:compacen}
\end{figure*}

\subsection{Atmospheric parameters}
\label{sect:result:parameters}

We measured the atmospheric parameters for the 985 spectra as indicated in Sect.~\ref{sect:tgm}.
As it is known from \citet{wu2011}, the automatic determination is 
highly reliable for the FGK stars, but lack of
faithfulness in some regions of the parameters' space. Namely, this
concerns the hot evolved stars and the
cool stars  ($T_{\rm{eff}} < 3600$~K). Therefore, for the stars
found in these regimes, we searched the literature for recent 
determinations based on high-resolution spectroscopy. 
We also examined the very low-metallicity stars, 
and those for which our derived parameters
depart significantly from those listed in \citet{cenarro2007}.
Whenever we found values judged more credible than ours, we adopted
them. 

For 77 stars (8\% of the library), we adopted parameters compiled and 
averaged from the literature.
For four of them
HD\,18191, 17491, 54810 and 113285, 
we adopted either the metallicity or the gravity from
the internal inversion of the MILES interpolator (see. Sect.~\ref{sect:result:inter}).
For six stars (one A-type star with emission line HD\,199478,
and five cool stars, $T_{\rm{eff}} < 3000$~K, G\,156-031 and 171-010,
HD\,113285, 126327 and 207076)
we could not
find any reliable source for at least one of the atmospheric parameters.

The most metal poor star of the library, HD~237846, belongs to a stream
discovered by \citet{helmi1999}. We adopted [Fe/H] $= -3.15$ from recent
measurements \citep{2009ApJ...706.1095Z,2010PASJ...62..143I,2010ApJ...711..573R},
while \citet{cenarro2007} catalogued [Fe/H]$ = -2.59$.
The inversion with ELODIE returned [Fe/H]$ = -2.52$.
The fitted  metallicity values for the low metallicity stars 
([Fe/H]$ < -1.70$) were often biased toward higher values by $\sim 0.15$ dex.
For 13 of these 46 metal deficient stars, we adopted parameters from the
recent literature

The adopted parameters are listed in Table~\ref{table:params},
also available in Vizier.
Figure~\ref{fig:distri_tg} shows the distribution of the stars in 
the $T_{\rm{eff}}$ vs. 
log~$g$ and  $T_{\rm{eff}}$ vs. [Fe/H] diagrams.

\setcounter{table}{1} 
\begin{table*}
\caption{\label{table:compa} Comparison of the atmospheric parameters with other studies.
}
\begin{tabular}{l|rrrrrrrr}
\hline\hline
Comparison & &N\tablefootmark{a} & \multicolumn{2}{c}{$T_{\rm{eff}}$} &  \multicolumn{2}{c}{log g (cm s$^{-2}$)} &  \multicolumn{2}{c}{[Fe/H] (dex)}\\
&&&\multicolumn{1}{c}{$\Delta$}&\multicolumn{1}{c}{$\sigma$}&\multicolumn{1}{c}{$\Delta$}&\multicolumn{1}{c}{$\sigma$}&\multicolumn{1}{c}{$\Delta$}&\multicolumn{1}{c}{$\sigma$}\\
\hline
\multirow{3}{*}{Cenarro} 
& OBA& 121&  2.1 \%&  7.9 \%& 0.080& 0.384& 0.101& 0.408\\
& FGK& 773&  46 K& 120 K& 0.038& 0.284& 0.045& 0.133\\
&   M&  91& -49 K& 165 K& 0.039& 0.317& 0.012& 0.283\\
\hline
\multirow{3}{*}{ELODIE} 
&OBA&  48& -3.2 \%&  4.7 \%& 0.026& 0.218& 0.009& 0.069\\
&FGK& 332&  12 K&  60 K& 0.008& 0.079& 0.038& 0.055\\
&M&  23&  -3 K&  16 K& -0.022& 0.200& 0.034& 0.061\\
\hline
\multirow{3}{*}{CFLIB} 
& OBA&  42& -2.2 \%&  6.0 \%&-0.016& 0.268& 0.025& 0.110\\
& FGK& 309&   2 K&  43 K&-0.025& 0.069& 0.021& 0.030\\
&  M&   16&  16 K&   9 K& 0.051& 0.173& 0.028& 0.061\\
\hline
\end{tabular}
\tablefoot{
For each parameter, the $\Delta$ column gives the mean difference `this work' $-$ `reference', and $\sigma$ the dispersion between the two series. The three lines are for
the OBA ($T_{\rm{eff}} > 8000 K$), FGK  ($4000 < T_{\rm{eff}} \leq 8000 K$) and M 
($T_{\rm{eff}} \leq 4000 K$) spectroscopic types respectively.
The statistics were computed with the IDL command \textsc{biweight\_mean},
to discard the outliers.\\
\tablefoottext{a}{Number of compared spectra}
}
\end{table*}

We compared, in Fig.~\ref{fig:compacen}, our parameters to those 
from \citet{cenarro2007}.
We also compared our results with ELODIE 3.2 and CFLIB \citep{wu2011},
for the stars in common,
and the corresponding statistics are shown in Table~\ref{table:compa}.

The mean deviations with \citeauthor{cenarro2007} are larger than those obtained
by \citet{wu2011} for the CFLIB library. For example, the dispersion
is 120~K for the FGK stars, while \citeauthor{wu2011} report 
dispersions of $\sim$~70~K when comparing
to homogeneous measurements based on high-resolution spectroscopy, and 
$\sim$~100~K when comparing to the compilation of  \cite{valdes2004}.
For the two other parameters, the dispersion is consistent with the comparison between
CFLIB and the \citet{valdes2004} compilation.
The comparisons with the ELODIE 3.2 and CFLIB parameters obtained with the same method, are
typical of comparisons between accurate spectroscopic measurements.

There is a statistically significant bias on $T_{\rm{eff}}$ of the 
FGK stars (47~K)
between our measurements and \citet{cenarro2007}. Although this is within the uncertainties
of the present calibrations, such a bias has consequences when the library is used
in models of stellar populations. As pointed in some
occasions \citep{prugniel2007, percival2009}, it is sufficient to alter the age derived for
old globular clusters by several Gyr.

We compared our measurements with \citet{gonzalez2009} who used the infrared flux method
to measure  $T_{\rm{eff}}$ for FGK stars using 2MASS photometry.
After clipping 9 outliers out of the 232 stars in common, we found that these values
are in average 28~K warmer than ours, with a dispersion of 141~K.
\citet{vazdekis2010} compared the \citet{cenarro2007} and \citet{gonzalez2009}
measurements and found a bias of 59~K of the same sign. 
Our measurements are in better agreement with \citet{gonzalez2009}
than the original MILES compilation, but
the different values of the bias are within the accuracy of the determination
of the temperature scale and are only marginally significant. 

We used the statistics of the comparison with \citeauthor{cenarro2007}
to estimate the external error.
We used the ratios of
the differences between the two series to the formal errors
to rescale the errors, conservatively assuming that the
mean precisions of each series are equivalent.
This rescaling factor depends on the temperature. It changes from 5 for the G stars
to about 20 for both the hottest and the coolest stars. These factors are the same
for the three parameters, and the same order of magnitude than those used in
\citet{wu2011}. The external errors are significantly larger than the formal
error for several reasons, including the internal degeneracies between the atmospheric
parameters. They are reported in Table~\ref{table:params}.

For the FGK stars the mean errors are 60 K, 0.13 and 0.05 dex respectively for 
$T_{\rm{eff}}$, log g and [Fe/H]. For the M stars, they are 38~K, 0.26 and 0.12 dex,
and for the OBA 3.5\%, 0.17 and 0.13 dex.
The figures are similar to the precision reported by \citet{wu2011}, implying
that there is no degradation of the performance of the method because of the
lower spectral resolution.

\subsection{Interpolator}
\label{sect:result:inter}

We adjusted an interpolator to all the stars in MILES, using the atmospheric
parameters of Table~\ref{table:params}.
For the 27 stars presenting a mean residual velocity shift greater than 30~\kms{},
we shifted the spectra by an integer number of pixels to reduce the effect. We
did not correct all the spectra for the wavelength dependent shifts derived in
Sect.~\ref{sect:result:lsf} to avoid a rebinning by fractions
of pixel.
We affected a weight to each star depending on its location in the parameters' space,
in order to compensate the uneven distribution of the stars. The low-metallicity stars,
and the coolest and hottest ones were over-weighted because they are in relative small
numbers. We did not weight with the signal-to-noise of the spectra because this information
is not available.

We checked the residuals between the observed and interpolated spectra to identify
and correct outliers. 
Finally, we assessed the validity of this interpolator performing two tests:
(i) We compared the original and interpolated spectra, and (ii) we used the
interpolator to measure the atmospheric parameters of MILES and CFLIB with 
ULySS.

\subsubsection{Detection and treatments of the outliers}

We started with all the stars, and
we examined the residuals between the observed and interpolated spectra.
There are a priori different causes for these residuals: (i) Although ``normal'' stars
were targeted, some peculiarities affect some of them (binarity, rotation, chromospheric 
emission, non-typical abundances ...), (ii) the atmospheric parameters
derived in Sect.~\ref{sect:result:parameters} have uncertainties (or errors), and (iii)
the MILES spectra have uncertainties.

The most prominent outliers correspond to spectra whose shape disagree with the
interpolator. This is probably not because of errors in the atmospheric parameters, 
as the spectral
features are generally well fitted, but rather because of errors in the flux calibration or
in the correction for Galactic extinction. 
We nevertheless searched the literature for indications of peculiarities that may
explain the discrepancies, and whenever we found some plausible reason we excluded
the star from the computation of the interpolator.
We observe that the spectra with wrong continuum shape are often located at low Galactic
latitude or in obscured regions. 
The most deviant example is HD\,18391, a Cepheid variable whose extinction was corrected 
assuming E(B-V) $=$ 0.205 mag. Our spectroscopic fit indicates a considerably
higher extinction, consistently with \citet{2009AN....330..807T} 
who derived E(B-V) $\approx$ 1 mag. Another example where the extinction was 
under-corrected by $\sim$~0.7 mag is HD\,219978. Although the main outliers
corresponds to underestimated extinctions, there are cases of over-estimation, 
like HD\,76813.

We suppose that the main source of discrepancy
is the correction of the Galactic extinction, but we cannot safely separate between this
possibility and an error on the flux calibration. 
Nevertheless, we assumed that for those discrepant cases, the error is due to the
extinction correction and we applied an additional correction using a Galactic extinction
curve \citep{schild1977}. Whenever this correction was unsatisfactory
(maybe because the source of error is the flux-calibration), 
we flagged the spectrum
to have a reduced weight or to be excluded.
We corrected the extinction for 55 field stars.

All this process was made iteratively, treating the most prominent outliers and 
recomputing a further version of the interpolator.
Finally, the mean residuals between the interpolated and observed spectra is 4\%, a value
comparable to what is obtained for the ELODIE interpolator. A large fraction of these
residuals are still due to mismatches of the shape of the continuum.

\begin{figure*}
\includegraphics{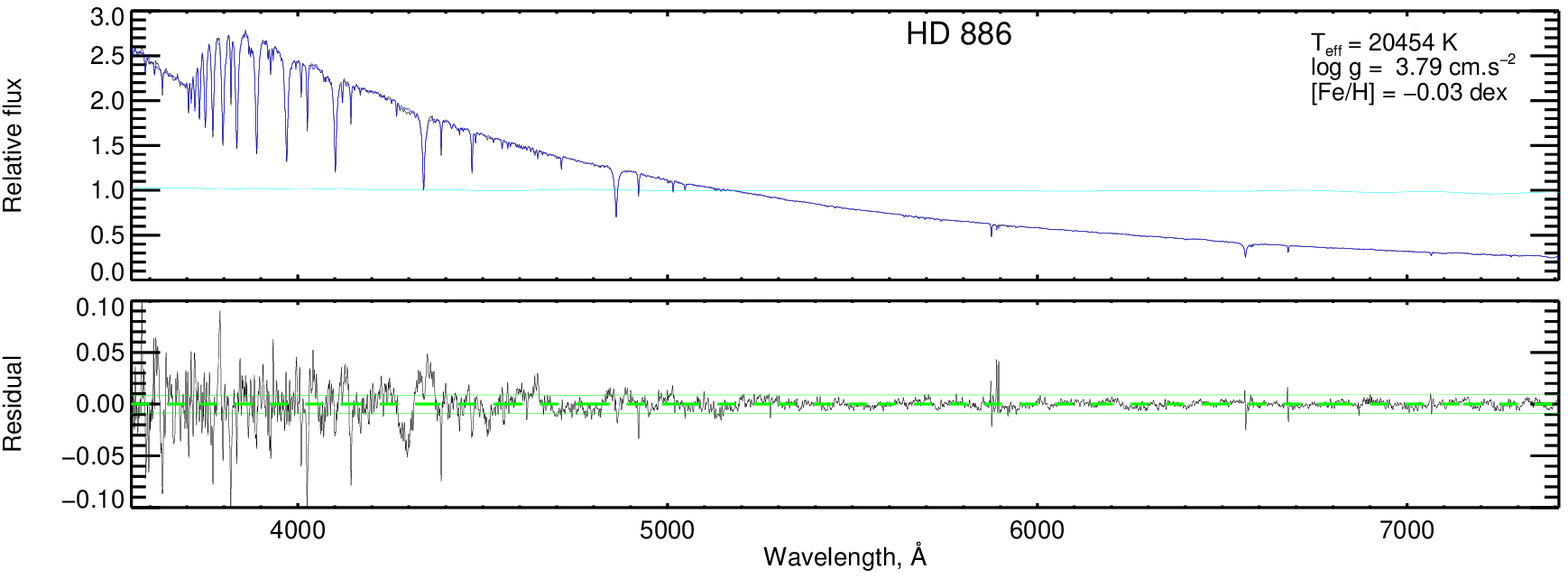} 
\includegraphics{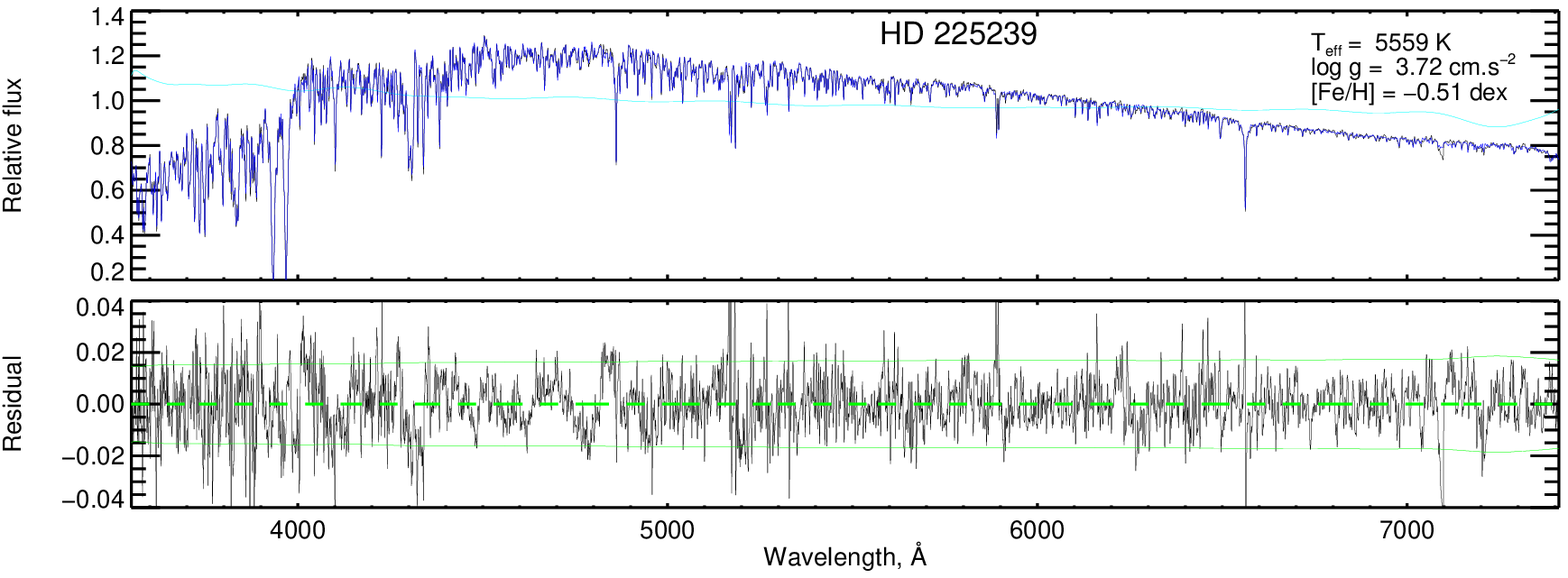} 
\includegraphics{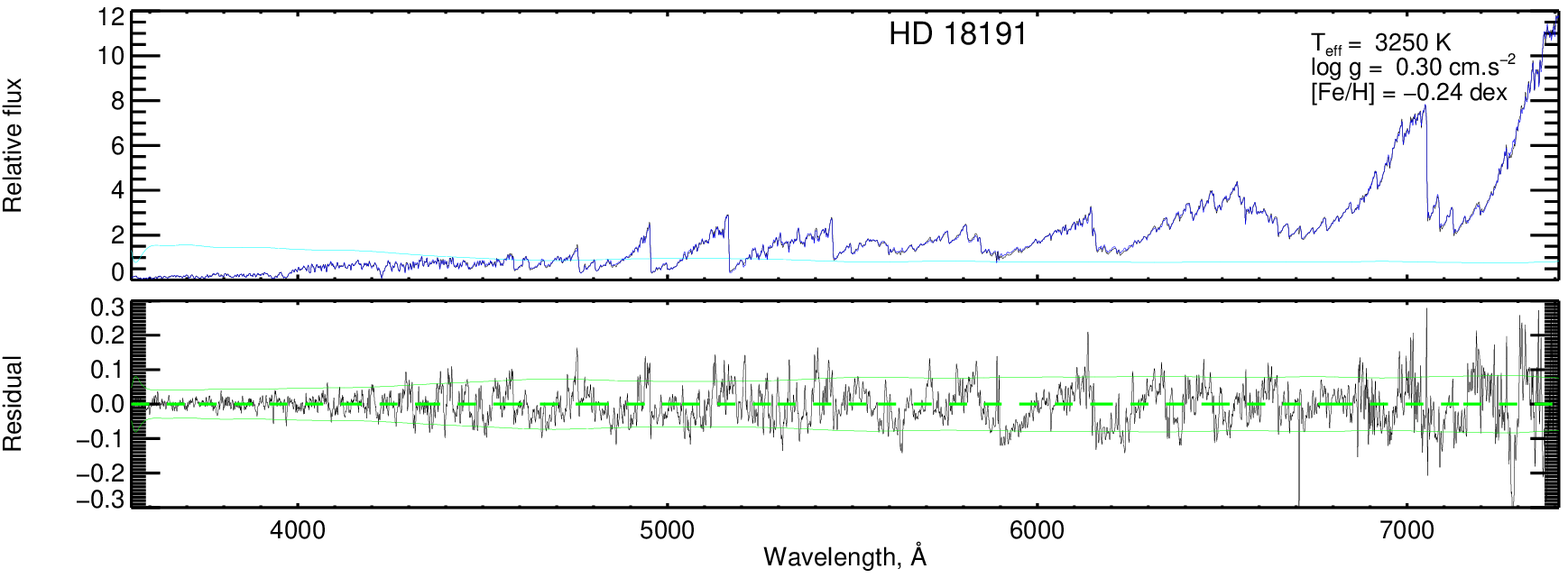} 
\caption {
Fits of MILES spectra with the MILES interpolator for three representative stars.
For each star, the top panel represents the flux distribution, normalized to an average
of one, and the bottom ones the residuals between the observation and the best fitted
interpolated spectrum (observation$-$model). The fit was performed with ULySS. 
The continuous green lines are the $\pm1 \sigma$ errors, assuming a constant error spectrum 
and $\chi^2 = 1$. 
The clear blue lines are the multiplicative polynomials.
}
\label{fig:fit}
\end{figure*}

\begin{table*}
\caption{\label{table:testinterp} Tests of the interpolator.
}
\begin{tabular}{l|rrrrrrrr}
\hline\hline
Comparison & &N\tablefootmark{c} & \multicolumn{2}{c}{$T_{\rm{eff}}$} &  \multicolumn{2}{c}{log g (cm s$^{-2}$)} &  \multicolumn{2}{c}{[Fe/H] (dex)}\\
&&&\multicolumn{1}{c}{$\Delta$}&\multicolumn{1}{c}{$\sigma$}&\multicolumn{1}{c}{$\Delta$}&\multicolumn{1}{c}{$\sigma$}&\multicolumn{1}{c}{$\Delta$}&\multicolumn{1}{c}{$\sigma$}\\
\hline
\multirow{3}{*}{MILES\tablefootmark{a}} 
&OBA& 130&  1.2 \%&  4.5 \%&-0.002& 0.202& 0.016& 0.118\\
&FGK& 770& -3.0 K& 68.7 K&-0.024& 0.108& 0.005& 0.074\\
&K5-M&  85& -7.7 K& 31.5 K& 0.011& 0.177& 0.039& 0.082\\
&M&  26&  2.1 K& 35.4 K& 0.092& 0.219& 0.045& 0.102\\
&BHB &  25&  2.7 \%&  9.1 \%&-0.095& 0.541&-0.092& 0.331\\
\hline
\multirow{3}{*}{CFLIB\tablefootmark{b}} 
&OBA& 231& -1.7\%&  6.3\%&-0.027& 0.214& 0.012& 0.157\\
&FGK& 960&-20.9 K& 75.4 K&-0.072& 0.104& 0.014& 0.063\\
&K5-M&  74&  6.4 K& 33.6 K& 0.069& 0.219& 0.089& 0.086\\
&M&  24& 31.8 K& 34.1 K& 0.163\tablefootmark{d}& 0.142& 0.179\tablefootmark{d}& 0.119\\
& BHB &  28& -6.9\%& 11.0\%&-0.243& 0.623&-0.707& 0.688\\
\hline
\end{tabular}
\tablefoot{
Comparison between the atmospheric parameters from this paper\tablefoottext{a}
and \citet{wu2011}\tablefoottext{b}  with those derived using the present MILES
interpolator.
For each parameter, the $\Delta$ column gives the mean difference `MILES interpolator' $-$ `reference', and $\sigma$ the dispersion between the two series. 
The different lines are for
the OBA ($T_{\rm{eff}} > 8000 K$), FGK  ($4000 < T_{\rm{eff}} \leq 8000 K$), K5-M 
($T_{\rm{eff}} \leq 4000 K$) and M 
($T_{\rm{eff}} \leq 3500 K$) spectroscopic types respectively. The BHB 
(blue horizontal branch) line is for the hot evolved stars with $T_{\rm{eff}} > 7000 K$
and [Fe/H]~$< -0.7$ dex.
The statistics were computed with the IDL command \textsc{biweight\_mean},
to discard the outliers.\\
\tablefoottext{c}{Number of compared spectra}\\
\tablefoottext{d}{The stars for which \citet{wu2011} give [Fe/H]$ = -1$ where rejected from the statistics on log g and [Fe/H].}
}
\end{table*}

\subsubsection{Tests of the interpolator}
\citet{wu2011} have shown that the ELODIE interpolator is not reliable for
the hot evolved stars nor for the very cool stars. It is not known if the
reason resides in the limited sampling of the parameters' space in these
regions or from more fundamental characteristics of the interpolator.
In order to check this, we used the new interpolator to measure the atmospheric
parameters of MILES and CFLIB.  
The first one is an internal test, where each MILES spectrum is compared to
the interpolator based on the whole library.
The statistics of the comparisons
between these new sets of parameters and the adopted ones are summarized
in Table~\ref{table:testinterp}.

For the coolest stars ($T_{\rm{eff}} \leq 3500 K$), 
the metallicities measured with the ELODIE interpolator were biased toward 
low values. This effect is absent with the MILES interpolator.
For the hot evolved stars ($T_{\rm{eff}} > 7000 K$ and [Fe/H]~$< -0.7$ dex)
the biases are also considerably reduced compared to those obtained with
the ELODIE interpolator.

Figure~\ref{fig:fit} presents the fits with the MILES interpolator of the MILES spectra
of three stars of different spectral types. The residuals are of the order of 1\%
of the flux, and the multiplicative polynomials are flat and close to unity, 
reflecting the good quality of the flux calibration of MILES.

These tests show that the MILES interpolator is reliable to measure the atmospheric
parameters over their whole range.
The lower resolution of MILES do not affect these determinations.
The FITS file containing the coefficients of the interpolator is available
in Vizier. It can be directly used in ULySS to fit
stellar spectra.
The interpolated spectra can also be computed online in a Virtual Observatory
compliant format \citep{prugniel2008}.

\subsection{Discussion on the flux-calibration}

The presumably good flux-calibration of MILES is its most attractive characteristics.
It was assessed by comparison with accurate broad-band photometry.
In order to test the photometric precision of the interpolator,
we fitted the residuals between the observed and
interpolated spectra with a straight line and expressed the result 
as a $\rm B-V$ colour.
For the whole library, we find $\rm \Delta(B-V) = 0.005$~mag 
and  $\rm \sigma(B-V) = 0.039$~mag, respectively
for the bias and dispersion. 
This residual colour is by construction small, since the interpolator was
built with the observed spectra, but the small dispersion reflects the good 
spectrophotometric precision. 
The photometric precision on the individual spectra were determined
by \cite{miles} to be $\rm \sigma(B-V) = 0.013 \sim 0.025$~mag, by comparing
synthetic $\rm B-V$ colours with different sets of standards
(the two numbers corresponds to different standards). Our present
values are not as precise, likely because they also include the errors on the
Galactic extinction corrections, on the atmospheric parameters and the 
cosmic variance introduced by characteristics of individual stars that
are not considered in the interpolator.

We can also assess the  photometric accuracy of the ELODIE library.
This has always been a question, because its flux-calibration results
from a complex and indirect process. 
To test it, we made series of interpolated ELODIE spectra following the
main and giant sequences and
we computed the photometric precision as above.
We found that the differences between the interpolated ELODIE and MILES 
spectra are $\rm \sigma(B-V) \approx 0.02$~mag, which is consistent
with the estimations made in \citep{elo30,elo31}.

\section{Summary and conclusion}
\label{sect:conclusion}

We derived the atmospheric parameters of the stars of the MILES library.
We estimated the external precision for the FGK stars to be 60 K, 0.13 and 0.05 dex 
respectively for $T_{\rm{eff}}$, log g and [Fe/H]. 
For the M stars, the mean errors are 38~K, 0.26 and 0.12 dex,
and for the OBA 3.5\%, 0.17 and 0.13 dex.
This precisions are comparable to those obtained with the same
method for the CFLIB library, whose resolution is significantly higher.
This shows that there is no significant degradation due to the resolution.

We characterized the LSF. We found that the residual shift of the rest-frame
reduction has a dispersion of 12~\kms, with an average of 2 \kms{}
(MILES is slightly red-shifted).
The mean FWHM dispersion of the library is 2.56~\AA, changing 
from 2.45 to 2.63~\AA{} from the blue to the red.

We computed an interpolator for the library. This is a function returning a
spectrum for given $T_{\rm{eff}}$, log g and [Fe/H].
In order to check its reliability, 
we used it to derive the atmospheric parameters of MILES itself and CFLIB.
The results are in good agreement with those derived
with the ELODIE interpolator in the present paper and in \citet{wu2011}.
For some regimes where the ELODIE interpolator has shown deficiencies
(hot evolve stars and cool stars), the MILES interpolator has better
performances.

In a companion paper, we will use this interpolator to prepare stellar
population models using PEGASE.HR.

\begin{acknowledgements}

We thank the referee for her/his constructive comments.
We acknowledge the support from the French 
{\it Programme National Cosmologie et Galaxies} (PNCG, CNRS).
MK has been supported by the Programa 
Nacional de Astronom\'{\i}a y Astrof\'{\i}sica of the Spanish Ministry 
of Science and Innovation under grant \emph{AYA2007-67752-C03-01}.
She thanks CRAL, Observatoire de Lyon, Universit\'{e} Claude
Bernard, Lyon 1, for an Invited Professorship.

\end{acknowledgements}

\bibliographystyle{aa} 
\bibliography{miles}  


\onltab{1}{
\longtab{1}{
\setlength{\tabcolsep}{4.5pt}
\begin{longtable}{lrrrrrrrrrl}
\caption{\label{table:params}Adopted atmospheric parameters for the 985 MILES stars}\\
\hline\hline
Name &
       Miles\tablefootmark{a} &
                 $T_{\rm eff}$ & error & $log(g)$ & error & [Fe/H] & error & 
        cz\tablefootmark{b} & $\sigma$\tablefootmark{c} & references\\
     &         & \multicolumn{2}{c}{(K)}            & \multicolumn{2}{c}{($cm\,s^{-2}$)} & \multicolumn{2}{c}{(dex)} & \\
\hline

\endfirsthead
\caption{continued.}\\
\hline\hline
Name &
       Miles\tablefootmark{a} &
                 $T_{\rm eff}$ & error & $log(g)$ & error & [Fe/H] & error & 
        cz\tablefootmark{b} & $\sigma$\tablefootmark{c} & references\\
     &         & \multicolumn{2}{c}{(K)}            & \multicolumn{2}{c}{($cm\,s^{-2}$)} & \multicolumn{2}{c}{(dex)} & \\
\hline
\endhead
\hline
\endfoot
\input parameters.tex
\hline
\end{longtable}
\tablefoot{
\tablefoottext{a}{Identification number in \citep{cenarro2007}.}\\
\tablefoottext{b}{Mean residual velocity shift resulting from the fit to the
          ELODIE interpolator. The shift may not be uniform throughout the 
          spectrum, because of uncertainties in the wavelength calibration
          of MILES.}\\
\tablefoottext{c}{Gaussian width of the absolute LSF at 5300 \AA.
          It includes the (variable) instrumental broadening and the physical
          broadening (rotation).}\\
\tablefoottext{d}{The references are coded as:
(0): This work; 
(0a): This work (interactive fit); 
(0b): This work (using the MILES interpolator); 
(1):  \citet{wu2011};
(2):  \citet{cenarro2007};
(3):  ELODIE 3.2, unpublished
(4):  \citet{2010AJ....140.1694F};
(5):  \citet{2010A&A...509A..20R};
(6):  \citet{2010A&A...515A..74L};
(7):  \citet{2010PASJ...62.1239T};
(8):  \citet{2010RMxAA..46..331A};
(9):  \citet{2009ApJ...702L..96G};
(10): \citet{2009A&A...501..297Z}; 
(11): \citet{2009BaltA..18...65B};
(12): \citet{2008AJ....136.1557H};
(13): \citet{2008A&A...478..507M};
(14): \citet{2008A&A...478..823M};
(15): \citet{2008A&A...481..777S};
(16): \citet{2008A&A...490..297S};
(17): \citet{2008MNRAS.383..729T}; 
(18): \citet{2008MNRAS.389..585C};
(19): \citet{2007BaltA..16..191K};
(20): \citet{2007MNRAS.378..617K};
(21): \citet{2006A&A...446..279C};
(22): \citet{2006A&A...453..895W};
(23): \citet{2005A&A...442..635B};
(24): \citet{2004A&A...418..989N};
(25): \citet{2002A&A...392.1031A};
(26): \citet{2001A&A...369..178B};
(27): \citet{2001A&A...380..630C};
(28): \citet{2001MNRAS.328..291T};
(29): \citet{2000A&A...353..978M};
(30): \citet{2000A&A...364..102K};
(31): \citet{1998MNRAS.296..856A};
(32): \citet{2009ApJ...692..522W};
(33): \citet{2010MNRAS.406..290G};
(34): \citet{2009PASJ...61.1165T};
(35): \citet{2008MNRAS.389.1336K};
(36): \citet{2009AN....330..807T};
(37): \citet{2005MNRAS.356..963W};
(38): \citet{2008MNRAS.390.1081H};
(39): \citet{2008ApJ...678.1329B};
(40): \citet{2010PASJ...62..143I};
(41): \citet{2009ApJ...706.1095Z};
(42): \citet{2010ApJ...711..573R};
(43): \citet{2003ApJ...588.1072B};
(44): \citet{2004ApJ...617.1091S};
(45): \citet{2007ApJ...660..747A};
(46): \citet{2004ApJ...607..474H};
(46): \citet{2009ApJ...701.1519R};
(47): \citet{2009A&A...497..611M};
(48): \citet{2008ApJ...681.1524L};
(49): \citet{2005A&A...430..507A};
(50): \citet{1999AJ....117..981B};
(51): \citet{2007A&A...475..519H};
(52): \citet{2008MNRAS.391...95R};
(53): \citet{2005MNRAS.364..712Z};
(54): \citet{2002AJ....123.1647S};
}
} 
} 
} 

\end{document}

%% file: parameters.tex
HD 224930              &0001& 5411&  36& 4.19&0.07&-0.78&0.04& -5& 59&0\\
HD 225212              &0002& 4179&  68& 0.85&0.19& 0.18&0.08& 21& 53&0\\
HD 225239              &0003& 5559&  44& 3.72&0.09&-0.51&0.05&-18& 59&0\\
HD 00004               &0004& 6779&  66& 3.87&0.07& 0.21&0.04&  1& 65&0\\
HD 00249               &0005& 4766&  53& 2.91&0.13&-0.27&0.06&  0& 58&0\\
HD 00319               &0006& 8641& 150& 4.29&0.09&-0.35&0.10&-18& 55&0\\
HD 00400               &0007& 6190&  51& 4.15&0.08&-0.22&0.04&-14& 63&0\\
HD 00245               &0008& 5749&  39& 4.13&0.07&-0.57&0.04&-14& 62&0\\